\documentclass[aps,pre,twocolumn,superscriptaddress]{revtex4} 
\usepackage{graphicx,epsfig}  
\usepackage{dcolumn}   
\usepackage{bm}        
\usepackage{amssymb}   
\usepackage{color}
\usepackage{subfig}
\usepackage{amsmath}
\usepackage{float}
\usepackage{caption}
\begin{document}

\title{Structure formation by  electrostatic interactions in strongly coupled medium}

\author{Mamta Yadav}
\email {ymamta358@gmail.com} 
\affiliation{Department of Physics, Indian Institute of Technology Delhi, Hauz Khas, New Delhi 110016, India}
		
\author{Priya Deshwal}
\affiliation{Department of Physics, Indian Institute of Technology Delhi, Hauz Khas, New Delhi 110016, India}
\author{Srimanta Maity}
\affiliation{ELI Beamlines Centre, Institute of Physics, Czech Academy of Sciences, Za Radnicí 835, 25241 Dolní Břežany, Czech Republic}
\author{Amita Das}
\email {amita.iitd@yahoo.com}
\affiliation{Department of Physics, Indian Institute of Technology Delhi, Hauz Khas, New Delhi 110016, India}

\begin{abstract}
The formation of correlated  structures is of importance in many diverse contexts such as strongly coupled plasmas, soft matter, and even biological mediums. In all these contexts  the  dynamics are mainly governed by  electrostatic interactions and  result in the formation of a variety of structures. In this   study,  the process of formation of  structures is investigated with the help of Molecular (MD) simulations in 2 and 3 dimensions.  The  overall medium has been  modelled with an equal  number of positive and negatively charged particles interacting via long-range pair Coulomb potential.  A repulsive short-range  Lennard-Jones (LJ) potential is added to take care of the blowing up of attractive Coulomb interaction between unlike charges.  In the strongly coupled regime, a variety of classical bound states form. However, complete  crystallization of the system,  as typically observed in the context of one component strongly coupled plasmas, does not occur.  
 The influence of localized perturbation in the system has also been studied.  The formation of a crystalline pattern of shielding clouds around this disturbance is observed.  The spatial properties of the shielding structure  have been analyzed using  the radial distribution function and Voronoi diagram. The process of accumulation of oppositely charged particles around the disturbance triggers a lot of dynamical activity in the bulk of the medium, wherein close encounters between  widely separated particles occur. This leads to  the formation of a larger number of bigger clusters. There are, however, also instances when bound pairs break up to provide the appropriate signed charge to the shielding cloud. A detailed discussion of these features has been provided in the manuscript. 

\end{abstract}

\maketitle
 \section{ Introduction}
Electrostatic interactions play an overwhelmingly important  role and  form the basis of many  processes, involving plasmas \cite{tsytovich2007development,filippov2015electrostatic,ratynskaia2006electrostatic}, soft matter \cite{naji2005electrostatic}, biological medium \cite{holm2001electrostatic,levin2002electrostatic}, etc.  In these cases, the system  is often modelled by variants of Langevin and Fokker Planck equations involving electrostatic interactions (e.g. the  Poisson Nernst Planck equations \cite{bazant2004diffuse,kilic2007steric}). Formations of colloidal suspensions, polymers, and clusters in these contexts have always attracted attention and often have significant technological and scientific relevance. Our attempt here is to understand  the process of structure formation with the help of Molecular Dynamics simulations in the simplest scenario of a collection of uniformly distributed charged particles of both signs, interacting  via  long-range Coulomb force along with Lennard Jones (LJ) interaction. The LJ potential helps in switching off the attractive Coulomb interaction amidst oppositely charged particles  at short distances preventing their collapse. One is thus seeking the formation of clusters in this medium, wherein two or more particles join together to form a stable bound  configuration. It is clear that the thermal effects will let the particles fly away inhibiting the survival of any such  binding configuration. However, if the medium is in a strongly coupled state for which the mutual interaction energy exceeds the thermal energy, the possibility of the formation of such a bound state exists. We, therefore, explore the dynamics in the strongly coupled regime for this medium. 
 
We have chosen  a collection of equal and oppositely charged particles interacting with electrostatic Coulomb and LJ potential in the strongly coupled regime for our model. This is essentially like a quasi-neutral strongly coupled plasma state of matter \cite{fortov2006physics,ichimaru1982strongly}. Preparation of a strongly coupled plasma  medium is usually a challenge as it  is typically created by doing violence to matter and hence is seething with thermal energy. The strongly coupled state requires that the parameter $\Gamma = Q^2/aK_bT > 1 $, where $Q$ is the charge state of the particle, $a$ is the  inter-particle separation and $T$ is the temperature and $K_b$ is the Boltzmann constant. Thus the strong coupling condition can be  achieved  by (i) enhancing the  charge $Q$, (ii) reducing the inter-particle distance $a$, and (iii) decreasing  the temperature $T$. 

Working with high $Q$ particles has led to the observation of strong coupling effects even at room temperatures in dusty plasmas  \cite{chu1994direct,thomas1994plasma} in the form of crystallization of highly charged micron-sized dust particles against a background of normal electron-ion plasma. The strongly coupled dusty plasma medium has garnered a lot of attention in the plasma community for a couple of decades now. The strong coupling effects playing a distinct role in the  collective modes have been studied extensively both theoretically \cite{kaw1998low} and experimentally \cite{murillo2004strongly}.  For the strongly coupled  dusty plasma medium properties like collective structures \cite{kumar2017observation,das2014collective}, linear and nonlinear waves \cite{kumar2018spiral}, instabilities \cite{tiwari2012kelvin,das2014suppression}, phase transitions \cite{maity2019molecular,maity2022parametric} have been observed. The interplay of single particle and collective dusty plasma  dynamics has also been studied \cite{maity2018interplay}
The dynamics of small dust clusters  have illustrated  dynamical equilibrium states with   chaotic motion \cite{deshwal2022chaotic,maity2020dynamical}. 
 
 In super-dense stars \cite{van1991dense}, the interior of planets,  inertial confinement plasmas, etc., the reduction of inter-particle separation $a$ can  lead to strong coupling behavior. In  some of these contexts, however, the dynamics might involve electromagnetic effects. 
There has also been tremendous progress in the creation of Ultracold plasma  experimentally using laser-cooling techniques. For instance,  xenon atoms via a photo-ionization process are confined by a magneto-optical trap \cite{killian1999creation,kulin2000plasma,killian2001formation} with typical density  in the range of $10^5 cm^{-3}$ to $10^{10} cm^{-3}$ with either both or one of the species  in a strongly coupled state. Many properties of ultracold plasma have been studied like the formation of Rydberg atoms \cite{killian2001formation}. Any heating in such a system needs to be avoided to keep the system in a strongly coupled state. For this, studies have been conducted to identify and restrict the inherent heating mechanisms that are prevalent in these systems.  Mechanisms such as  disorder-induced heating \cite{cummings2005fluorescence,guo2010molecular} and three-body recombination(3BR) \cite{kuzmin2002numerical} have been identified as some prominent factors causing the heating of the medium. A study of thermodynamics and transport properties such as pressure and internal energy \cite{tiwari2017thermodynamic} etc., has also attracted attention. 
  It is also interesting to note that  studies in magnetically  confined plasma  \cite{gilbert1988shell} have shown ions  in the strong coupling regime arranged in  concentric shells and demonstrate solid and liquid-like behavior. In another  work by T. Pohl et. al., \cite{pohl2004coulomb} the crystallized structure was observed at the center of expanding laser-cooled ultracold plasma in which short-range concentric ion shells are formed with appropriate initial conditions. 
   
   Our objective here is to explore the possibility and nature of structure formation in a neutral plasma. The study also has relevance in other contexts of soft matter, and biological systems in which structure formation is an important issue, and electrostatic interactions are  believed to be responsible for the same. Here, one is seeking the formation of structures  with the most elementary  coulomb interaction with Lennard Jones potential at short distances to avoid the collapse of unlike charged particles. We  make use of the open-source molecular dynamics code of LAMMPS for carrying out these studies numerically. 
    
  This article has been organized as follows. In Section II, we  discuss the simulation setup and the choice of  parameters.  In  section III, we investigate the  formation of  classical bound states. In sections  IV and V, we study the effect of an external perturbation introduced in the 2-dimensional  and 3-dimensional  simulation  systems. The perturbation is in the form of inserting a massive and heavily charged  particle into the medium. This could be equivalent to applying a biased probe in experiments. This inserted particle remains  fixed in space while the particles in the medium respond to its potential.  The charge in it is  chosen to be quite high,  about two orders higher than the charge of other particles in the system. The evolution of the system is studied in both  two and three  dimensions.  Oppositely charged particles assemble around the externally placed charge particle to screen its potential. However, when the underlying plasma is strongly coupled, the oppositely charged particles assemble in a  crystalline pattern around it. The behavior of the crystalline pattern is understood with the help of  the Radial distribution function and Voronoi diagrams. As expected  the potential due to this crystalline charge cloud  typically resembles  the  Debye shielding profile. However, there are some deviations,  arising  due to the crystalline and discrete nature of the shielding cloud. The value  of $\sigma$, which typically defines the distance where the  LJ repulsive potential goes to zero, is also found to play a role. The role of external perturbation on the formation and existence of the seemingly fragile classical bound state is studied.    It is observed that the disturbance induced by the insertion of the external highly charged particle  leads to the formation of higher  numbers of  bigger-sized bound state clusters. This happens as the particles placed far apart  now get an opportunity for close encounters. However, in all these simulations  no three-dimensional bound state structure formation was observed. In section VI it is shown that  when some two-dimensional bound states are evolved in isolation (absence of other plasma particles) they  reorganize as three-dimensional bound structures. In fact, it is shown that the 3-d structures that form are  energetically favorable. Yet such structures do not form in the backdrop of  plasma medium. It clearly shows that the plasma medium plays an important role in the formation of the clusters. 
  In Section VII we summarize and conclude our studies. 

  \label{intro}

\section{ MD Simulation Details}
\label{mdsim}

We have carried out 2D and 3D Molecular Dynamics simulations using an open-source classical MD simulation software LAMMPS \cite{plimpton1995fast}. We have typically used the parameters associated with electron-ion plasma in an ultra-cold regime. The particles in the simulation box is interacting with each other through the long-range pair Coulomb potential ($V_{pC}$) and an additional short-range repulsive LJ potential ($V_{LJ}$) as given by   
\begin{equation}
V(r_{kl}) = V_{pC}+ V_{LJ}
\label{total_potential}
\end{equation}
where,
\begin{equation}
{V}_{pC}= \frac{Q_{k}Q_{l}}{4\pi\epsilon_{0}r_{kl}}
\label{Coulomb_potential}
\end{equation}

\begin{equation}
{V}_{LJ}= 4\epsilon\left[\left(\frac{\sigma}{r_{kl}}\right)^{12}-\left(\frac{\sigma}{r_{kl}}\right)^{6}\right]
\label{lj_potential}
\end{equation} 
 This short range LJ potential avoids the recombination of electrons and ions. Here $Q_k$ and $Q_l$ define the $k^{th}$ and $l^{th}$ particles respectively.  Also,  $r_{kl}$ represents the distance between these two  particles. Here, $\epsilon$ and $\sigma$ are the  usual Lennard-Jones potential parameters where $\epsilon$ defines the strength of the LJ potential well and $\sigma$ describes the distance at which inter-particle LJ potential becomes zero. The PPPM (particle-particle particle-mesh) method \cite{heath1997proceedings} is used for the calculation of long-range Coulomb interactions. In our case, we have assumed both electrons and ions to be in a strongly coupled regime which means the coulomb coupling parameter ($\Gamma$) for both species is greater than unity. 

 In the two-dimensional (2D) study, we have chosen a rectangular simulation box of length $L_x = L_y = 1.4\times10^{-3}$m in the $\hat x$ and $\hat y$ directions, respectively. Periodic boundary conditions are considered in all directions. Initially, $10^{4}$ electrons and $10^{4}$ ions with densities $1.0 \times 10^{10}$ m$^{-2}$ are distributed randomly inside the simulation box. This ensures that there exist no initial spatial pre-correlations between the charged particles. The charge $(Q_i)$ and mass $(m_i)$ of the ion are taken to be $1.6\times 10^{-19}$ C and $100 m_e$, respectively. Here, $m_e$ is the mass of an electron, i.e., $m_e = 9.109 \times 10^{-31}$Kg. The mass of the ion in simulation is chosen  smaller compared to even the proton mass, merely to reduce the simulation time. In our simulations, the temperature of electrons and ions are both considered to be equal and at $T_e = T_i = 0.1$ K. All the length scales are normalized by the average inter-particle distance $a$. In 2D, the average inter-particle distance $a$ and the areal density $n$ are related as $a = (n\pi)^{-1/2}$. Thus for our choice  of $n$, the value of average inter-particle distance turns out to be  $a = 5.64\times10^{-6}$m. In our simulations, the time scales are normalized by $\omega_{pe}^{-1}$, where $\omega_{pe}=(ne^2/m_e\epsilon_{o})^{1/2}$ is electron-plasma frequency. 

 In our 3D simulations, we have taken the length of the simulation box to be $L_x = L_y= L_z= 3.7\times 10^{-5}$ $m$, in $\hat x$, $\hat y$, and $\hat z$ directions, respectively. Again, $10^{4}$ electrons and $10^{4}$ ions with densities $3.7 \times 10^{17}$ m$^{-3}$ are distributed randomly inside the simulation box. The relation between density $n$ and average inter-particle distance is $a = (3/(4n\pi))^{1/3}$. The temperature of electrons and ions is chosen to be  $T_e = T_i = 1$ K,  which corresponds to the Coulomb coupling parameter $\Gamma = 20$ i.e. in the strongly coupled regime.

 As mentioned earlier, particles (both, electrons and ions) are interacting with long-range pair Coulomb potential. Additionally, a short-range LJ pair potential is present in the pair interactions between particles. The parameters associated with these interaction potentials are as follows. The cutoff distance for Coulomb pair potential is chosen  to be $r_c = 20a$. The parameter $\sigma$ associated with the LJ potential is chosen to be $10^{-2}a$ for most of the simulation studies. We have, however, also varied the value of $\sigma$.  The normalized parameter $\epsilon$ is chosen to be $558 K_BT$ and $75 K_BT$ for 2D and 3D, respectively. 
  
 In both 2D and 3D simulations, Nose-Hoover thermostats \cite{nose1984unified,hoover1985canonical} have been used to achieve the thermal equilibrium associated with the desired value of Coulomb coupling parameter $\Gamma$. 

\section{ Formation of bound structures for the system in equilibrium} 
\begin{figure}[hbt!]
 \includegraphics[height = 9cm,width = 9cm]{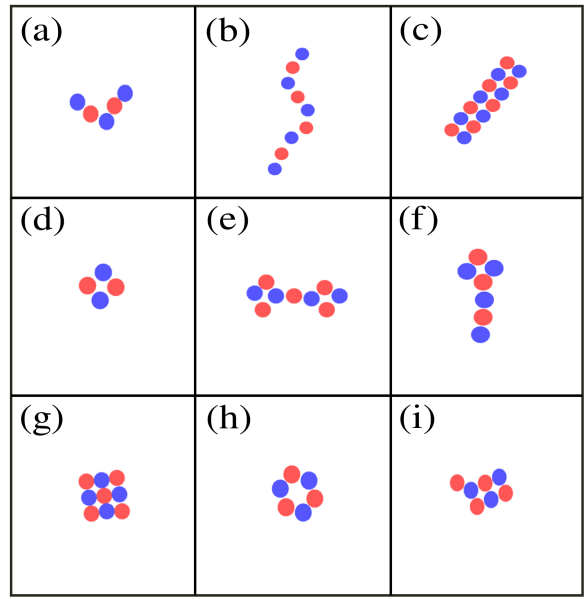}
     \caption{Distinct variety of bound structures at equilibrium. Here, the red and blue dots corresponds to the electrons and ions, respectively.}
  \label{Fig:bound-structures}
 \end{figure}
 
Once the system achieves thermodynamic equilibrium with the attached thermostat,   its  evolution shows the formation of several distinct varieties of bound clusters. In these structures, a few  electrons and ions combine with each other and form  a stable configuration. Some such structures  have been displayed in Fig.\ref{Fig:bound-structures}. There are linear structures as shown in the subplot (a) and (b) of this figure. In fact in subplot (b) about 9 particles (4 electrons and 5 ions) have joined together to form a linear chain. Such linear chains have alternating particles with opposite signs attached together. Subplot(c) shows a 2-D structure in which  two linear chains  join adjacently to form a 2-d structure. In subplot(d) we have shown the formation of a square structure in which the diagonally opposite particles are of the same sign. Two such square structures can join together as shown in subplot(e) or a square can have a linear chain attached to itself as shown in subplot(f) resembling a kite. Subplot (g) is again a square structure with $9$ particles, each side of the square having $3$ particles. The circular ringed structure in subplot (h) forms when the particles at the end of the linear chain join together. Subplot (i) is again a kite-like structure like that shown in (f). It should be noted that the bound structures that form do not necessarily have an equal number of ions and electrons. A mismatch of one unit charge (but not higher than this)  has been observed.  
 \begin{figure}[hbt!]
   \includegraphics[height = 9cm,width = 8cm]{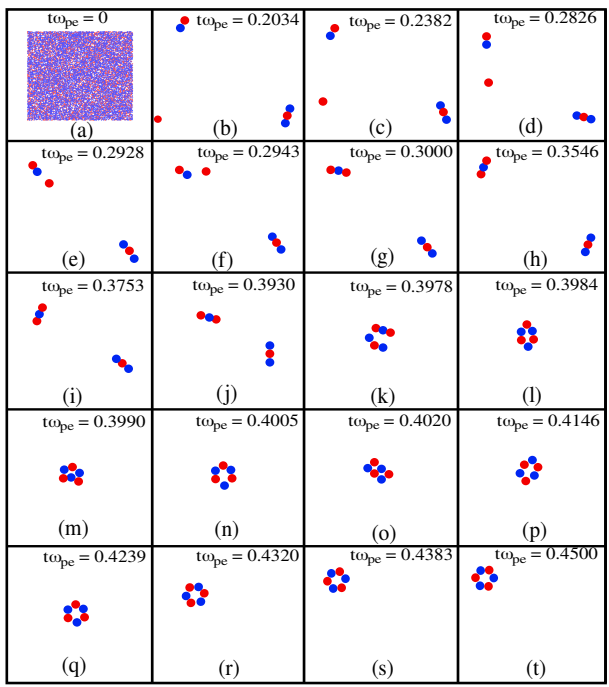}
     \caption{The schematics for the time evolution of one kind of bound structure. Blue dots and red dots represents the ions and electrons. The normalized time is varying from $t\omega_{pe}$ $0$ to $0.45$.}
  \label{Fig:dynamical-formation}
 \end{figure}

We illustrate the dynamical process of formation of some of these structures in Fig.(\ref{Fig:dynamical-formation}) by showing the snapshot of a particular zoomed region from the entire simulation region which has been shown in the subplot(a) of the figure at $t = 0$, where the randomly distributed particles have been shown. From subplot (b) to (t) we show the snapshots at various times for a zoomed region from the simulation box containing $6$ particles. The entire dynamics is covered within a fraction of  the plasma period. During this period no other particle entered this region hence the particle number remained at $6$.  From subplot (b) to (f) one can observe the attachment of an electron to an ion-electron pair placed at the top of the box. A $3$ particle linear chain can be observed to be rotating at the bottom right corner of the subplot. From subplot(g) to (k) one can observe how the two $3$ particle linear chains approach each other and join together. After combining, they wiggle a lot going through a variety of phases shown in subplots (l) to (p). Finally, as they relax they  form a ringed structure which is found to be stable.  

 We would like to emphasize here that, while the simulations have been carried out both in 2-D and 3-D, we observe no formation of three-dimensional clusters. 
 In the next section, we observe what happens to these structures when the medium is perturbed by inserting a highly charged heavy particle having no dynamics of its own. We also study how the plasma particles accumulated around it to shield the potential of this externally placed particle in the medium. 
\section{Structure formation in a two-dimensional perturbed system}  
When a weakly coupled  plasma is disturbed by an external potential, the plasma tries to shield this potential within a distance of Debye length.  It is, however, not clear how a strongly coupled plasma would react to an externally applied field. In this section, we try to explore this particular issue in a two-dimensional simulation set-up.  Once the system gets settled in thermal equilibrium an external point charge is  inserted at the center of the simulation box. The mass ($M_{P}$) and charge ($Q_{P}$) of external perturbation are chosen to be large as $100 m_i$ and $100 Q_i$, respectively. Since the externally inserted charge has a very high mass it remains stationary and is not allowed to move. The electrons get attracted towards this externally applied positive charge whilst the ions are repelled from it. The insertion thus initiates a dynamical response from the plasma. We investigate the behavior of the shielding cloud that accumulates around this charge  and also the influence of such a dynamic response from the plasma on the classical bound pairs.  It is observed that the shielding occurs in the form of an electron cloud arranged in a crystalline pattern due to the strongly coupled nature. It is interesting to note that despite the plasma getting  perturbed, even bigger-sized bound state clusters form in the bulk region. We discuss these two effects in the subsequent subsections. 

\subsection{The structure of shielding cloud}

\begin{figure}[hbt!]
   \includegraphics[height = 6.5cm,width = 7.5 cm]{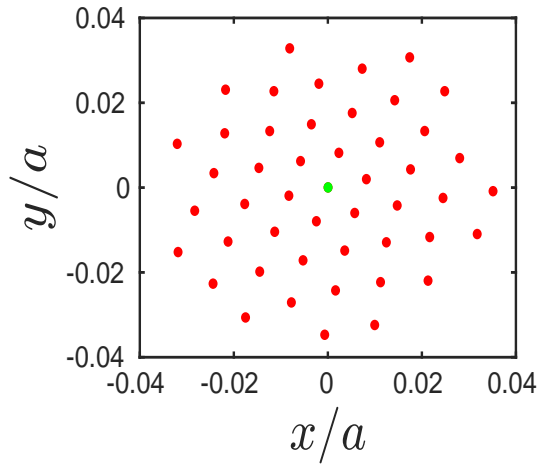}
    \caption{ Arrangements of electrons around external perturbation. Green dot represents the external perturbation that is introduce at the center of the simulation box and red dots represents electrons.}
  \label{Fig:shielding-cloud}
\end{figure}

\begin{figure}[hbt!]
   \includegraphics[height = 7.5cm,width = 9.0cm]{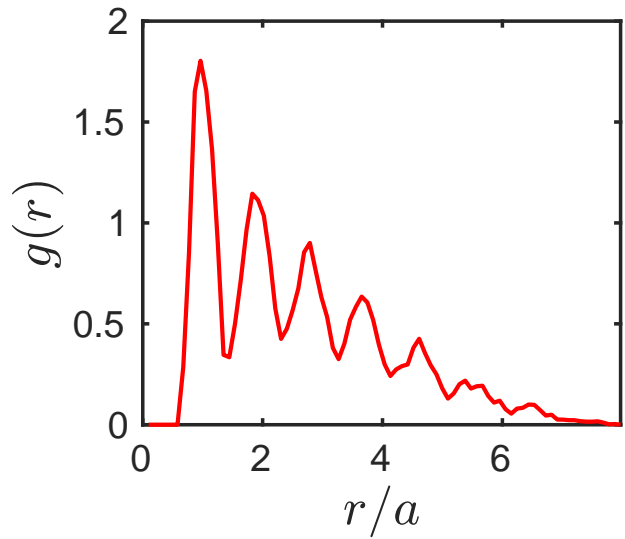}
     \caption{Radial distribution function as a function of distance from external perturbation at time ($t\omega_{pe}$) $15.93$}
  \label{Fig:rdf-2D}
 \end{figure}

Fig.\ref{Fig:shielding-cloud} represents the arrangements of electrons around the external perturbation at a time ($t\omega_{pe}=15.93$) after the charge has been inserted. The central green dot represents the externally inserted high positively charged perturbation, and the red dots correspond to the  electrons that accumulate around it. The process of accumulation is slow as electrons came one by one from bulk plasma  and get arranged in a hexagonal lattice structure of Fig.\ref{Fig:shielding-cloud}. After completing the first shell electrons start arranging in the next shell and so on. 
We show the radial distribution function (RDF) measuring the probability distribution of finding the electrons with the inserted external particle as the reference in Fig.\ref{Fig:rdf-2D}. The various peaks in the structure correspond to the various rings/shells that  are present in the crystalline structure formed by the electrons around the reference particle that has been introduced. 

\begin{figure}[hbt!]
   \includegraphics[height = 7.5cm,width = 9.0cm]{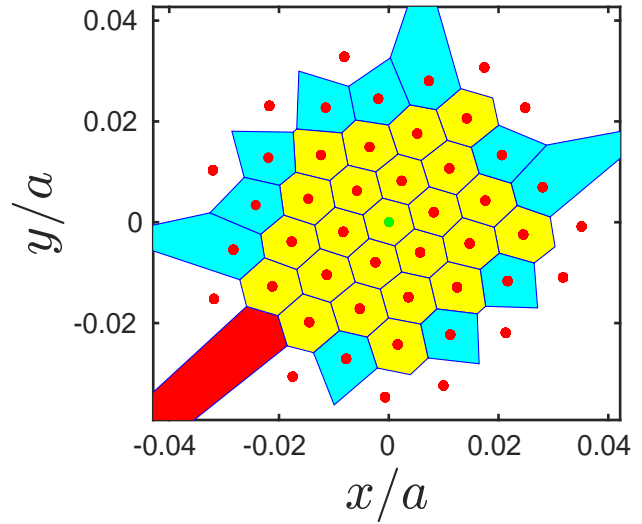}
     \caption{Voronoi diagram of electrons around external perturbation. Voronoi cells in yellow color represents six nearest neighbor whereas, cyan color denotes five neighbors. }
  \label{Fig:voronoi-2D}
 \end{figure}
 Fig.\ref{Fig:voronoi-2D} represents the Voronoi diagram for the shielding cloud  Fig.\ref{Fig:shielding-cloud}. This provides information about  the nearest neighbour of each particle in the cluster by constructing polygons around each point. The boundary of the polygon is constructed by bifurcating the distance between the particle and its nearest neighbours. We illustrate the formation of hexagons and pentagons  with yellow and cyan colours, respectively. It is clear from the Voronoi diagram that the shielding cloud arranges itself in the hexagonal lattice. The pentagonal structure arises at the boundary of the cloud.  

\begin{figure}[hbt!]
   \includegraphics[height = 7.5cm,width = 9cm]{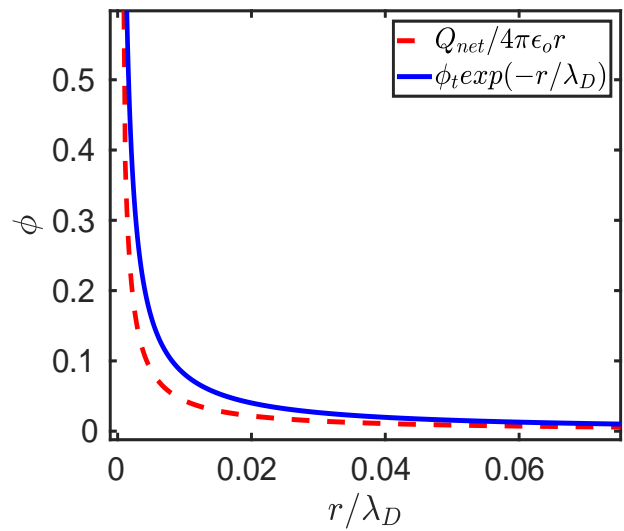}
     \caption{Plot of potential profile as a function of normalized distance from external perturbation having  charge ${100 Q_i}$. Here blue and red curve corresponds to potential in normal electron-ion plasma and ultracold plasma, respectively. }
  \label{Fig:potential-2D}
 \end{figure}

 Once the shielding process gets  completed and the plasma becomes quiescent, we decide to study the potential profile of the shielding cloud to determine whether it resembles the Debye shielding form encountered in the weakly coupled regime. The value of  Debye length as estimated from the choice of our simulation parameters  is ${\lambda_D} = 1.7\times10^{-4} m$. The theoretical Debye shielding profile using this particular value of $\lambda_D $ along with the one obtained from the simulation has been shown in Fig.\ref{Fig:potential-2D} by solid blue and red dashed lines, respectively. The plot clearly shows that while there is an agreement in both the plots at higher values of $r/\lambda_D$, at shorter distances the  simulation depicts a sharper screening. Here, the value of $\sigma = 5.64\times10^{-8}m$ is chosen to be  shorter than the Debye length. In fact we observe that the value of $\sigma$ plays a crucial role in defining the screening now. In the next section, this is discussed in the context of 3-D simulations. 

\begin{figure}[hbt!]
   \includegraphics[height = 8.5cm,width = 6.0cm]{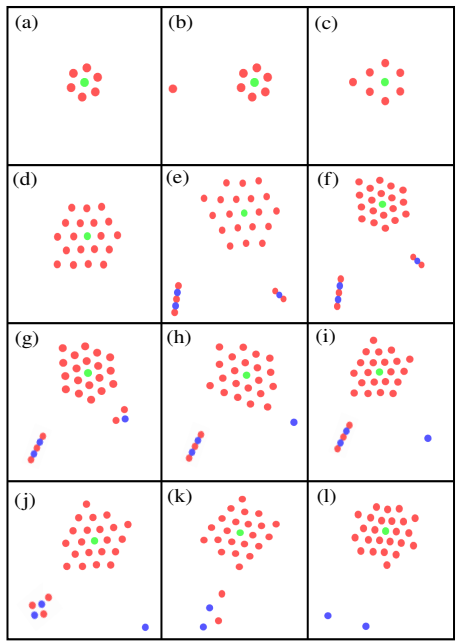}
     \caption{Evolution of structure around the external perturbation. Here, red and green dots corresponds to electrons and externally introduced perturbation,respectively. }
  \label{Fig:structure_evolution_2D}
 \end{figure}

\subsection{Impact on bound structures in bulk}

\begin{figure}[hbt!]
   \includegraphics[height = 9cm,width = 7cm]{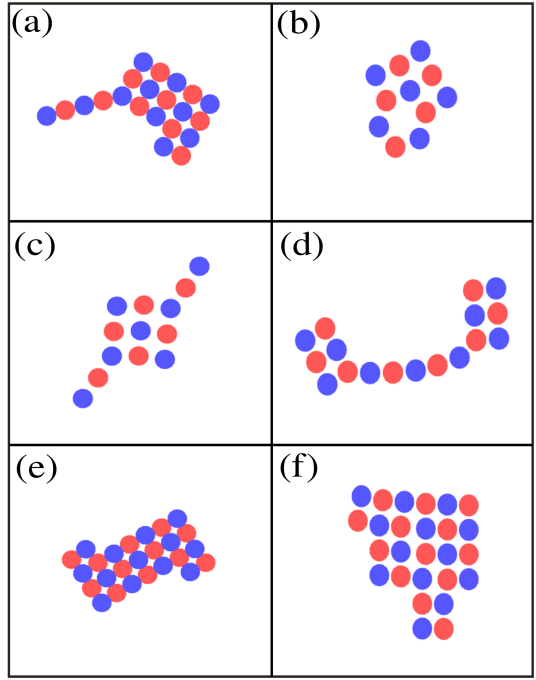}
     \caption{More complex bound structures formed in bulk of plasma when the external perturbation is introduced.}
  \label{Fig:perturbed-bound}
 \end{figure}

We now study the effect this perturbation has on the formation of  bound structures in the bulk. The perturbed system  initiates a dynamical response by the plasma. While the electrons get preferentially attracted to shield the charge of the particle introduced in the medium, they even drag along with them some of the bound structures to which they are attached.  In Fig.(\ref{Fig:structure_evolution_2D}), we show through snapshots of various times,  how free electrons get attracted towards the disturbance one by one and get attached to it forming a lattice structure around it. It can be observed from subplots $(e)$ to $(l)$, that even bound states get dragged by the electrons toward the structure. Upon reaching a very close distance the bound cluster gets dissociated and leaves the  electron in it to contribute to the screening cloud, while the ion rebounds back to the bulk. However, despite such disassociation of the bound structures,  it is interesting to note that the increased activity  allows for more encounters amidst  particles in the bulk resulting in larger and more complex shaped structure formation as has been depicted in Fig.(\ref{Fig:perturbed-bound}). These bigger structures were not observed in simulations when no external perturbation was introduced. In fact, in Fig.\ref{Fig:structure_evolution_2D} itself, it can be observed that a linear chain hovering around the structure at the left-down corner gets converted into a complex kite-like pattern in subplot (j). However, subsequently, it too being very close to perturbation donates all its electrons to the shielding cloud and the two ions bounce back. 
\section{External perturbation in 3D system} 
 We have also carried out our simulation studies in three dimensions. We distribute  $10^{4}$ electrons and $10^{4}$ ions  randomly  in the simulation box and seek the attainment of thermodynamic equilibrium at the desired temperature. Thereafter, we perturb  the system by inserting  a massive particle having a  charge of  $100 Q_i$  at the center of the simulation box. Since the inserted particle has a positive charge again the electrons in the system rush to shield its charge. This time, however, the system is three-dimensional so the electrons form a three-dimensional shielding cloud. We now investigate whether the shielding cloud has a crystalline form and if so what kind of the arrangement takes place in the cloud. As the structure cannot be visualised in 3-D we have employed certain diagnostics to perceive it in the best possible manner.
\subsection{Form of the shielding cloud}
\begin{figure}[hbt!]
   \includegraphics[height = 4.5cm,width = 8.8cm]{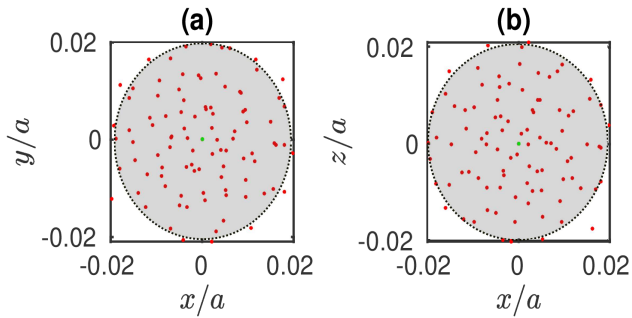}
     \caption{The projection of all the electrons that are involved in structure formation around external perturbation (green dot) in (a) horizontal plane (x-y plane) and (b) vertical plane (x-z plane).}
  \label{Fig:projection-3D}
 \end{figure}
 
\begin{figure}[hbt!]
   \includegraphics[height = 7.5cm,width = 9.0cm]{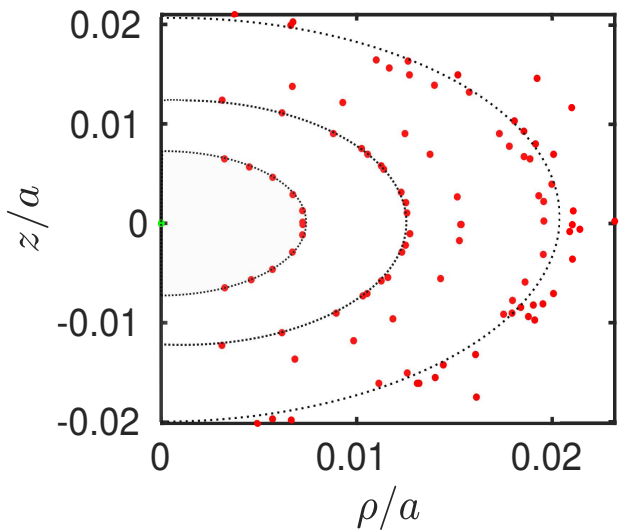}
     \caption{Shell structure by the projection of all the electrons into the $\rho-z$ plane. Most occupied Shells are represented by the black dotted lines and the green dot corresponds to external perturbation.}
  \label{Fig:shellstructure-3D}
 \end{figure}

\begin{figure}[hbt!]
   \includegraphics[height = 8cm,width = 8cm]{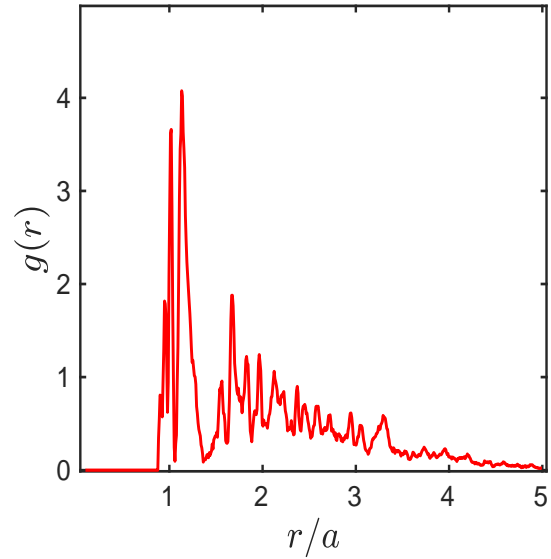}
     \caption{Radial distribution function as a function of distance from the externally introduced perturbation at time ($t\omega_{pe}$) $218.7$.}
  \label{Fig:rdf-3D}
 \end{figure}

 In Fig.\ref{Fig:projection-3D} subplots (a) and (b) the projection of all electrons in the $x-y$  and  $x- z$ planes have been shown.  The central green dot  is the reference point for the  highly charged particle introduced as an external perturbation in the system.  In both subplots, almost all positions are enclosed within the dotted circles which signifies that the cluster structure has isotropic spherical confinement. The arrangement appears to be random and isotropic from this perspective.  However, the shell structure of the arrangement is pretty evident when we choose to plot the locations of the  electrons involved in shielding in terms of cylindrical coordinates of $\rho$ ( $\rho = \sqrt{(x^2+y^2)}$) and $z$ in  Fig.\ref{Fig:shellstructure-3D}. As $z$ could be any axis so it is evident that the arrangement is in the form of spherical shells around the external perturbation introduced in the medium. 
The radial distribution function plotted in Fig.\ref{Fig:rdf-3D} having peaks confirms further the formation of the shell structure. 

\begin{figure}[hbt!]
   \includegraphics[height = 9cm,width = 8cm]{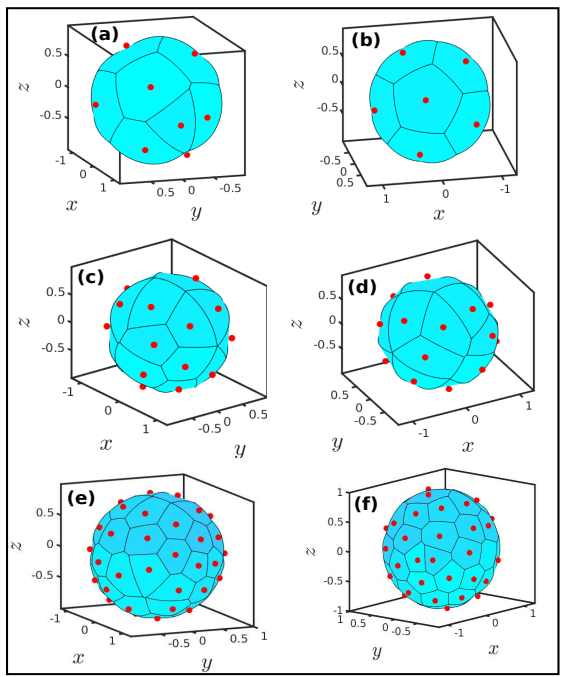}
     \caption{Voronoi diagram with two different orientations for electrons in  First shell (shown in subplot (a)-(b)), second shell (shown in subplot (c)-(d) ) and in third shell (shown in subplot (e)-(f)) having 12,20 and 50 number of electrons, respectively.}
  \label{Fig:voronoi-3D}
 \end{figure}

 The Voronoi diagram in the 3-D spherical shell has been shown in Fig.{\ref{Fig:voronoi-3D}} at the radius of the various rings determined from the plot of Fig.{\ref{Fig:shellstructure-3D}}. The coordinate axes 
 depicted  in Fig.{\ref{Fig:voronoi-3D}} are   normalised by the radius of that particular  shell. The number of polygons appearing in various shells has been counted and shown in Table. {\ref{table}}. The pentagons appear to be the dominant form in the first two shells. However, the third shell shows hexagons to be dominant. With different choices of charges taken for the  inserted particles, we observe the same  spherical arrangement of electrons. The innermost shell has 12 electrons and the second and the last shell have 20 and 50 electrons, respectively. 
 
  \begin{table}[hbt!]
\centering
  \caption{Number of polygons}
  \begin{tabular}{|c|c|c|c|c|c|}
  \hline
       Shell &Octagon & Heptagon & Hexagons & Pentagons & Tetragons \\ \hline
       
        First &0 & 0 & 1 & 9 & 0 \\ \hline
        Second &0 & 2 & 6 & 8 & 1 \\ \hline
        Third &1 & 4 & 27 & 17 & 0 \\ \hline
   \end{tabular}
   \label{table}
 \end{table}

 \subsection{Time evolution of structure in 3D} 
 We also investigate how the shielding process progresses in time. For this purpose, we track the number of electrons as they collect around the externally introduced particle.  
    \begin{figure}[hbt!]
   \includegraphics[height = 6.5cm,width = 9.0cm]{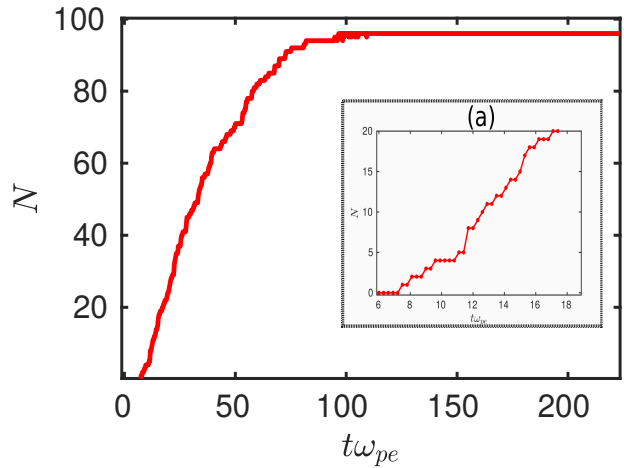}
     \caption{ Plot of number of electrons (${N}$), near the external perturbation, with normalised time. Subplot (a) corresponds to zoomed plot at initial time.}
  \label{Fig:time-evolution-3D}
 \end{figure}
 In  Fig.\ref{Fig:time-evolution-3D} we show the number of electrons (${N}$) which have been collected as a function of  the time of their arrival in units of  $\omega_{pe}$. It is clear from the figure that it takes  $100s$ of plasma period for the electrons to arrive and shield the charge almost completely. While initially, the accumulation is relatively fast, it slows down considerably at the later stage. There is no change after about $100 \omega_{pe}$ 
 when about $96$ electrons have accumulated in the shielding cloud. 
 
\subsection{ Comparison with Debye shielding}
\begin{figure}[hbt!]
   \includegraphics[height = 7.5cm,width = 9.0cm]{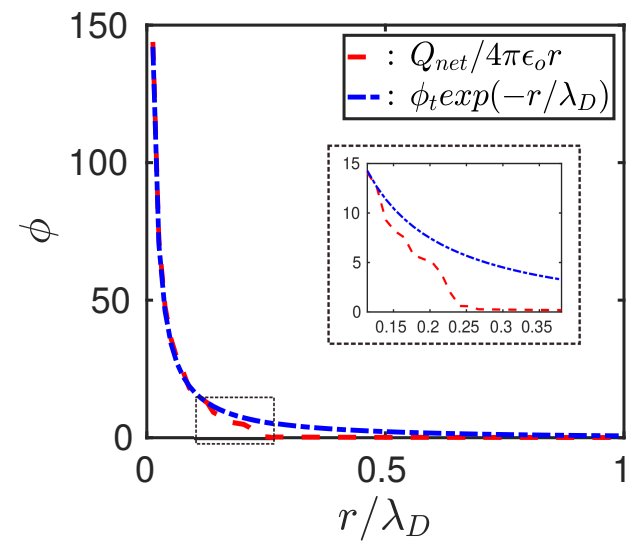}
    \caption{ Plot of potential profile as a function of normalized distance from external perturbation having charge ${100 Q_i}$ and value of LJ parameter ($\sigma$) is $8.64\times10^{-9}$ i.e. smaller than the Debye length ($\lambda_{D}$) $7.88\times10^{-8}$. Here, blue and red curve corresponds to potential in normal electron-ion plasma and ultracold plasma, respectively and subplot (a) represents the zoomed plot of the rectangular region shown by dotted lines.}
  \label{Fig:potential-3D}
\end{figure}

We now study the 3-D radial potential profile obtained numerically after the shielding process has taken place and compare it with the Debye shielding profile. Such a comparison  has been shown in Fig.(\ref{Fig:potential-3D}). The blue solid line shows the theoretical Debye shielding profile whereas, the red dotted line is the potential profile that has been obtained from the simulation. Here too like 2-D, we observe that the numerically obtained potential profile differs from  the theoretical Debye profile. In fact, due to the discrete arrangement of the electrons in various spherical shells, a sudden fall is also noticed in the profile. The zoomed-inset shows this  more clearly in the figure.  The location of this fall does not seem to be of statistical nature and occurs at the same location for the chosen set of parameters in a particular simulation confirming the uniqueness of the 3-D pattern that forms. 

\begin{figure}[hbt!]
   \includegraphics[height = 7.5cm,width = 9.0cm]{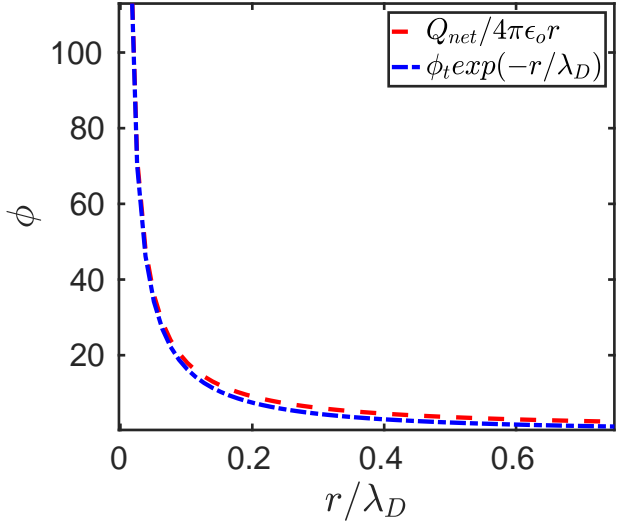}
     \caption{Plot of potential profile as a function of normalized distance from external perturbation. Here, value of LJ parameter ($\sigma$) is $8.64\times10^{-8}$ i.e. larger than the Debye length ($\lambda_D$). Blue and red curve corresponds to potential in normal electron-ion plasma and ultracold plasma, respectively.}
  \label{Fig:potential-sigmachange}
 \end{figure}

We have also simulated the 2D and 3D ultracold electron-ion plasma with different values of LJ parameters i.e. epsilon (${\epsilon}$) and sigma (${\sigma}$). We observed that even when $\epsilon$ and $\sigma$ are changed the shielding electrons still accumulate along various spherical shells. However, the location of the sharp dip in the potential profile does  change when the value of $\sigma$ in the Lenard Jones potential is varied. When we increase the value of parameter $\sigma$ to such an extent  that it is greater than the Debye length ($\lambda_D$) the shielding of the potential becomes broader.  This  has been illustrated in Fig.(\ref{Fig:potential-sigmachange}). 

\section{3D bound structures}
In the simulations that we have carried out so far, we have observed only 2-D bound state formation. The bound state shows considerable stability and survives for a long time in the plasma where it faces constant bombardments from other particles. In this section, we explore the stability of a couple of chosen bound structures in isolation by tracking their evolution. 
 
\begin{figure}[hbt!]
   \includegraphics[height = 8.5cm,width = 6.5cm]{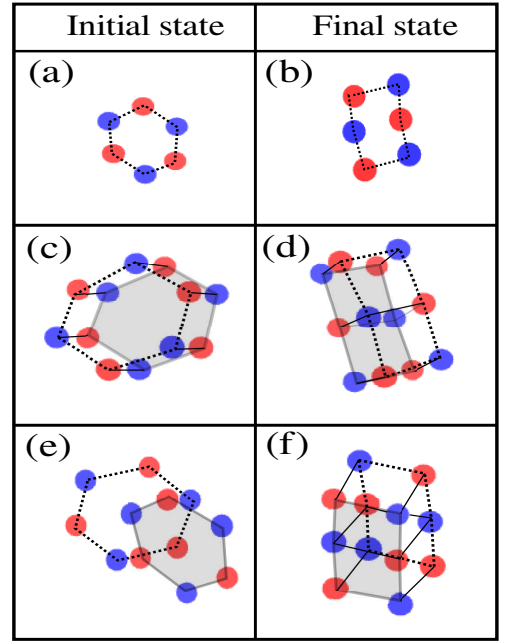}
     \caption{Schematic representation of initial and final states of bound structures in 3D (without background plasma). Here, red and blue dots corresponds to electron and ions and subplots (a),(c) and (e) represents the initial states whereas, subplots (b), (d) and (f) shows the final states.}
  \label{Fig:3d-boundstates}
 \end{figure}

   We first study the stability of the ring structure shown in subplot (a) of Fig({\ref{Fig:3d-boundstates}}) as an initial configuration  and track its evolution. It quickly (within a fraction of the plasma period) attains the form shown in subplot(b). We also placed a pair of rings on top of each other  (subplot(c) ) and a linked pair of rings (subplot(e). Both these  configurations relax immediately to a cuboid form as shown in subplots (d) and (f) of Fig(\ref{Fig:3d-boundstates}). It appears as a puzzle that the ring structure which was stable in the presence of background plasma and survive for several plasma periods, is unstable and quickly relax to the cuboid structure in the absence of background plasma. It should also be noted that the final configuration is a 3-D bound state which was never observed in the plasma. 
 
\begin{figure}[hbt!]
   \includegraphics[height = 4.0cm,width =8cm]{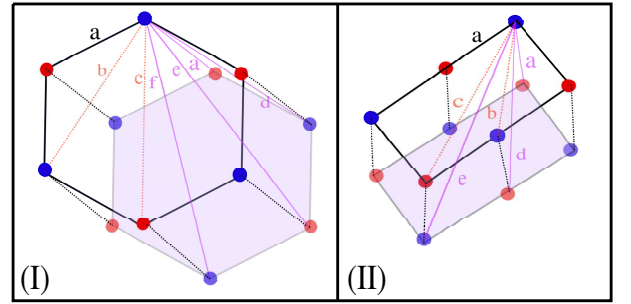}
     \caption{Schematic representation of ringed and the cuboid structure. }
  \label{Fig:ring_cuboid}
 \end{figure}

We now look at the energetics of the ringed and the cuboid structure.   
The interparticle spacings of the two-ringed and the cuboid structure are shown in Fig(\ref{Fig:ring_cuboid}). The total potential energy of the ringed structure and the cuboid structure in terms of the interparticle spacings shown in Fig(\ref{Fig:ring_cuboid}) has been evaluated. 
 
       \begin{equation}
           PE_{pC}^{ring}= 6\frac{Q_i Q_j}{4\pi\epsilon_0}\left(\frac{-3}{a}+\frac{2}{b}-\frac{1}{c}+\frac{2}{d}-\frac{2}{e}+\frac{1}{f}\right)
       \end{equation}
       \begin{equation}
           PE_{lj}^{ring}=6(2V_{lj}(a)+2V_{lj}(b)+V_{lj}(c)+2V_{lj}(d)+V_{lj}(e)+V_{lj}(f))
       \end{equation}
       whereas, all the parameters in terms of $a$ are
        \begin{center}
             $b=2a sin(60^{\circ})$, $c= 2a$
      
       $d=\sqrt{2}a$,  $e=2a$, $f=\sqrt{5}a$
     
    \end{center}
    The total potential energy of ringed structure is
    \begin{center}
      $  PE^{ring}(a) = PE_{pC}^{ring}+PE_{lj}^{ring}$
    \end{center}

 whereas, 
 \begin{equation}
           PE_{pC}^{cuboid}= \frac{Q_i Q_j}{4\pi\epsilon_0}\left(\frac{-18}{a}+\frac{22}{b}-\frac{8}{c}-\frac{8}{d}+\frac{4}{e}\right)
       \end{equation}
      
 \begin{equation}
 \begin{split}
 PE_{lj}^{cuboid}=20V_{lj}(a)+4V_{lj}(2a)+22V_{lj}(b)+8V_{lj}(c)+ \\
 8V_{lj}(d)+4V_{lj}(e)
\end{split}
 \end{equation}
 whereas, all the parameters in terms of $a$ are

 \begin{center}
       $b=\sqrt2a$,    $c=\sqrt5a$   
     
       $d=\sqrt6a$, $e=\sqrt3a$
 \end{center}
 The total potential energy for cuboid structure is
  \begin{center}
      $  PE^{cuboid}(a) = PE_{pC}^{cuboid}+PE_{lj}^{cuboid}$
    \end{center}

 \begin{figure}[hbt!]
   \includegraphics[height = 7.5cm,width = 9cm]{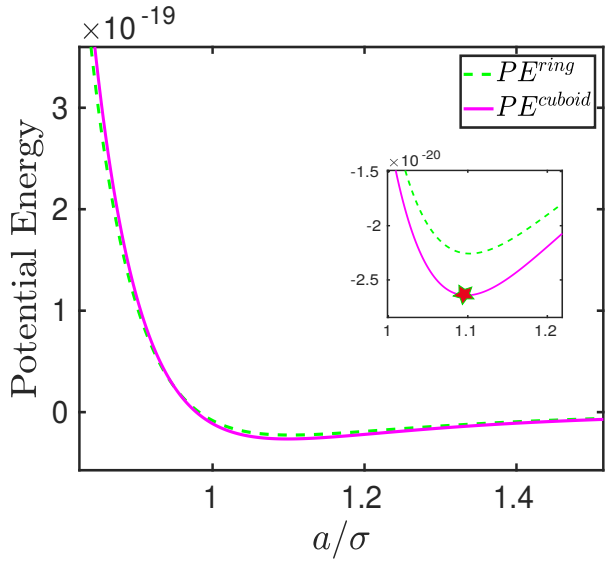}
     \caption{Plot of total potential energy of ringed (green) and cuboid  structure (magenta) as a function of interparticle spacing which is normalized with the LJ parameter $\sigma$. Here, star represents the minima in the potential energy and zoomed view of minima is shown in inset.}
  \label{Fig:potential_bound}
 \end{figure}
 We have then plotted the expression for the potential energy for the two cases as a function of $a/\sigma$ as shown in  Fig.{\ref{Fig:potential_bound}}. The green and magenta curve corresponds to overlapping ring and cuboid structure, respectively. The potential energy of the cuboid structure is found to be less than the overlapped ring structure. We believe this is the reason for favoring their formation in isolation. The presence of plasma clearly seems to modify this. 
 

 \section{summary} 

We have simulated an ultracold collection of an equal number of negative and positively charged particles (e.g. like electrons and ions, albeit their mass ratio is chosen to be different from the realistic case for ease of computation) in 2D and 3D by MD simulation in which both species are strongly coupled. These electrons and ions-like particles are interacting with long-range Coulomb potential, and a short-range LJ pair potential is present in the pair interaction between particles.  At equilibrium, we have observed many types of bound structure formation. With the introduction of external perturbation (highly charged and massive point particle)  we observe the expected phenomena of shielding. The particles are in a strongly coupled regime, therefore,  the shielding cloud arranges in specific patterns. 
The shielding potential profile unlike the weakly coupled cases is determined by the choice of the LJ parameter of $\sigma$ defining the distance at which the interparticle potential is minimum. In the crystalline pattern, the inter-particle distance gets determined by this parameter. Thus when $\sigma $ is larger than the Debye length, the shielding is broader than expected from the Debye screening process. On the other hand, when it is smaller than the Debye length the screening is sharper. Thus one needs to be cautious while drawing inferences  of certain  plasma properties while simulating the Ultra Cold strongly coupled plasma using  the short-range repulsive form to overcome the blowing up of the attractive potential amidst oppositely charged particles.

 \begin{center}
    \textbf{ACKNOWLEDGEMENTS} 
 \end{center}

  This research work was supported by a J.C. Bose fellowship grant (Grant No. JCB/2017/000055/SSC) from the Department of Science and Technology (DST), Government of India and the Core Research Grant (Grant No. CRG/2018/000624) of the science and Engineering Research Board (SERB), Government of India. The authors thank IIT Delhi HPC facility for computational resources. M.Y. is thankful to the University Grants Commission [Grant No. 1316/CSIR-UGC NET DEC.2017] for funding the research.

\bibliography{cluster_ref}
\end{document}